\documentclass[sigconf]{acmart}
\AtBeginDocument{%
  }
    
\setcopyright{none}
\acmDOI{}
\acmISBN{}

\acmConference[LIMITS '26]{12th Workshop on Computing within Limits}{June 23--25, 2026}{Online}

\begin{document}

\title{Addressing Negative Commons Governance with Positive Commons Principles}

\author{Boyang Zhou}
\authornote{Both authors contributed equally to this research.}
\email{zby2003@cs.washington.edu}
\orcid{0009-0004-1774-4345}
\affiliation{%
  \institution{University of Washington}
  \city{Seattle}
  \state{Washington}
  \country{USA}
}

\author{Oleg Ianchenko}
\authornotemark[1]
\email{olegian@cs.washington.edu}
\orcid{0009-0003-1077-3939}
\affiliation{%
  \institution{University of Washington}
  \city{Seattle}
  \state{Washington}
  \country{USA}
}

\renewcommand{\shortauthors}{Boyang Zhou, Oleg Ianchenko}

\begin{abstract}

    Computing is accompanied by both positive and negative commons throughout its lifecycle of creation, execution, and disposal.
    We examine two governance systems situated within this lifecycle -- global e-waste trade and the Linux kernel community -- to evaluate whether Elinor Ostrom's eight design principles for common-pool resource (CPR) governance extend to the management of negative common-pool resources (NCPRs).
    Unlike traditional CPRs where communities work to preserve a finite resource (i.e. clean water), NCPR governance seeks to collectively reduce a negative shared stock.
    In our two cases, e-waste governance aims to reduce the volume of mismanaged waste and illicit trade, while the Linux community aims to reduce the number of error-prone or malicious contributions that reach the main branch and, in turn, extend the life of existing hardware.
    Through qualitative analysis of primary sources from each domain, we find that the same eight principles by Ostrom that aid positive commons governance tend to appear in successful negative commons governance systems.
    We argue that future NCPR governance design should prioritize Ostrom's principles, particularly clearly defined boundaries and well-functioning nested structures.
\end{abstract}



\keywords{Negative common pool resources, Linux kernel, E-waste, Basel Convention, Governance}


\maketitle

\section{Introduction} \label{sec:intro}

The \emph{tragedy of the commons} was formalized in 1968 by ecologist Garrett Hardin, in an essay of the same name \cite{hardin_tragedy_1968}.
Hardin's argument leans on the observed over-exploitation of resources which are held in common, in other words, resources in arenas where it is difficult -- if not impossible -- to restrict public access to them.
More generally, the tragedy relates to the managerial pitfall of any publicly held stock when appropriator's self-interested decisions are used to control the overall exploitation of the resource.
He details the over-grazing of public land as an example: the owner of a herd has an individual incentive to utilize as much of the grazable land as possible to maximize their personal benefit. However, this individual incentive is at odds with the shared, collective incentive to maintain the land's viability for future grazing.

In the same essay, Hardin describes a different kind of commons: pollution.
Unlike a grazing field where actors extract value, here actors add to a shared burden.
In this kind of situation, polluters have an individual incentive to dump as much waste into a river as possible to avoid the high costs of proper disposal.
In tension with this individual incentive, these same actors nonetheless share a collective incentive to keep the river clean to avoid detrimental environmental effects.
Hardin notes the inverse relationship between these two problems: while appropriators of a positive commons share an incentive to maintain the resource stock, contributors to a polluted river share an incentive to reduce or eliminate the stock entirely.

Building on Hardin's initial observations, contemporary scholars like Mies and Monnin have formalized and extended the definition of what constitutes a negative commons \cite{monnin_negative_2021, mies_defending_2001}.
They expand this concept to include not just environmental pollution, but a broader class of shared realities and detritus that nobody wants, such as polluted soils, abandoned infrastructure, or the rejects of the technosphere, which nonetheless require active, collective stewardship \cite{monnin_capitalisme_2023, monnin_politiser_2023}.
Throughout this paper, we use the term Negative Common Pool Resources (NCPRs) to describe these situations where a community must govern and reduce a shared negative stock \cite{monnin_negative_2021}.
By establishing this distinction -- where positive commons require managing extraction and negative commons require managing contribution -- we can better analyze the institutional structures needed to govern them.

When looking for successful models of collective governance, the foundational framework remains the work of Elinor Ostrom.
In her Nobel-winning work, Ostrom conducted careful empirical work on successfully managed positive CPRs around the world \cite{ostrom_governing_2015}.
She compares CPRs in Japan, Switzerland, the Philippines, Spain, Sri Lanka, T\"{u}rkiye, and many other communities, deriving a list of eight key principles (further described in \S\ref{sec:eight-pds}) that lead to successful communal management of the pool.
Ostrom's findings ultimately prompted Garrett Hardin to revise his earlier analysis.
As commons scholarship notes, Hardin's initial model lacked empirical evidence and confused jointly managed commons with open-access regimes—a flaw he formally acknowledged in 1998 by narrowing his focus to the "unmanaged commons" \cite{frischmannRetrospectivesTragedyCommons2019a}.

While Ostrom's principles were explicitly derived to preserve positive commons, the fundamental hurdle of collective action -- aligning individual incentives with shared community goals -- remains identical for NCPRs.
Therefore, we hypothesize that the same institutional architectures that prevent the over-extraction of a positive resource might also prevent the over-pollution of a negative one.

To examine this hypothesis, we ground our analysis within the world of computing.
While common-pool problems involving positive resources often present as purely technical challenges, like efficient job scheduling or bandwidth allocation, the dilemmas surrounding NCPRs tend to stem from complex institutional challenges.
To understand these challenges, we look across the entire lifecycle of computing, from the digital infrastructure that powers fundamental systems to the physical hardware that inevitably becomes obsolete.

To that end, we selected two seemingly disparate case studies that represent the extreme ends of this lifecycle: the open-source Linux kernel and the global governance of electronic waste (e-waste).
We chose these specific examples as they perfectly illustrate the dichotomy between immaterial and material NCPRs.
The Linux community governs an immaterial digital commons, tasked with preventing the accumulation of error-prone or malicious code from degrading a shared system.
Conversely, e-waste represents a massive, material NCPR, where international frameworks like the Basel Convention attempt to manage the physical detritus of discarded hardware and prevent the externalization of toxic disposal costs.

Taking inspiration from Ostrom's empirical approach to positive commons, we evaluate these two computing NCPRs to answer the following research questions:
\begin{itemize}
    \item[\textit{RQ1:}] To what extent does the institutional design of an NCPR governance structure adhere to Ostrom's eight design principles for CPRs?
    \item[\textit{RQ2:}] To what extent do local manifestations of the NCPR governance structure adhere to Ostrom's eight design principles for CPRs?
    \item[\textit{RQ3:}] Broadly, do Ostrom's eight principles contribute to successful management of NCPRs?
\end{itemize}

\section{Background}
We begin by giving a description of the taxonomies and classifications used throughout this paper, followed by a brief description of the two NCPR governance structures we analyze.

\subsection{Ostrom's Eight Principles} \label{sec:eight-pds}
Ostrom's pivotal work, \emph{Governing the Commons} \cite{ostrom_governing_2015}, performs a thorough analysis of successful governance structures of CPRs around the world.
She distills out the following eight guiding principles:
\begin{enumerate}
    \item \textbf{Clearly defined boundaries:} The extents for who can appropriate from -- or contribute to -- a CPR must be well determined. This includes clear definitions of the resource itself, its boundaries, and the conditions under which participation is permitted.
    \item \textbf{Congruence between appropriation/provision rules and local conditions:} Rules related to how much or how often a resource is extracted from the pool must be related to the local conditions of the pool itself.
    \item \textbf{Collective-choice arrangements:} Rules governing the commons must be set by those who the rules affect.
    \item \textbf{Monitoring:} Appropriator behaviors and the state of the resource pool must be monitored.
    \item \textbf{Graduated sanctions:} Violations of rules must be met with sanctions, the seriousness of which must depend on the seriousness of the infraction.
    \item \textbf{Conflict-resolution mechanisms:} An accessible method for resolving issues between appropriators, monitors, or otherwise any participants in the commons must exist.
    \item \textbf{Minimal recognition of rights to organize:} Appropriators must be able to create their own governance structures; these local structures must be recognized by all higher-level governance structures.
    \item \textbf{Nested enterprises:} CPRs that are part of larger systems must be organized into multiple layers. This goes hand in hand with the right of appropriators to organize.
\end{enumerate}

\subsection{Negative Commons Taxonomy}

The term "negative commons" was originally coined by Mies and Bennholdt-Thomsen to describe organic waste that once circulated locally but became unmanageable under mechanized, privatized agriculture \cite{mies_defending_2001}.
As introduced previously, Monnin substantially broadened this concept to encompass the persistent physical and epistemic legacies of industrial capitalism. 
He characterizes these not as mere byproducts, but as \emph{ruins}: residues of obsolete systems that resist decay and continue to impose costs long after the systems that produced them have ended \cite{monnin_negative_2021, monnin_capitalisme_2023, monnin_politiser_2023}.

To classify these shared, unwanted stocks, two complementary frameworks inform our analysis.
First, Deron \cite{deron_sprouting_2025} builds directly on Monnin to propose a taxonomy oriented around the role an NCPR plays in a sustainability transition. 
He outlines four categories: \emph{ruins} (material elements such as obsolete infrastructure that persist past their usefulness), \emph{ghosts} (the immaterial paradigms, expectations, and beliefs that licensed those ruins), \emph{seeds} (material elements that could seed a more sustainable future, such as localized repair networks), and \emph{visions} (the immaterial counterparts that orient those seeds).

The main distinction running through Deron's scheme is the distinction between \emph{material} and \emph{immaterial} elements.
Because our aim is to test whether Ostrom's principles carry across fundamentally different types of NCPRs, we collapse Deron's four-way scheme to this single dichotomy. 
Selecting e-waste as a material NCPR and the Linux kernel as an immaterial NCPR lets us set aside the forward-looking \emph{seeds} and \emph{visions} and focus strictly on the governance of existing negative stocks (the \emph{ruins} and \emph{ghosts}).

Second, an alternative classification by Monnin organizes negative commons by \emph{what should be done with them}: those we must learn to live with, those we must live with differently, and those we must live without \cite{monnin_negative_2021}.
This action-oriented distinction is independent of the material/immaterial divide and bears directly on how we interpret our findings; we therefore return to it in our discussion (\S\ref{sec:limitations}).

\subsection{E-waste}
E-waste, or Waste Electrical and Electronic Equipment (WEEE), is recognized as one of the fastest-growing solid waste streams in the world.
In 2022 alone, an estimated 62 million tons of e-waste were produced globally, a figure projected to rise significantly in the coming decade \cite{GlobalEwasteMonitor2024}.
This waste stream is physically complex and comprises over 60 different elements from the periodic table.
It contains both valuable resources, including gold, copper, silver, and critical raw materials like cobalt and indium, alongside a cocktail of hazardous materials such as lead and mercury.

The processing of this waste often occurs outside of formal industrial facilities.
A significant portion of global e-waste is handled through "informal recycling", where rudimentary methods are used to extract valuable metals.
These methods include open burning of electrical cables to recover copper, and acid leaching to extract gold from circuit boards \cite{GlobalEwasteMonitor2024}.
These processing practices release heavy metals and organic pollutants directly into the surrounding environment (soil, air, water), exposing workers and entire communities to severe health risks.

\subsubsection{E-waste and Negative Commons}
E-waste is a clear example of material NCPR.
As with other pollution-based negative commons problems, actors have incentives to externalize disposal costs due to the high cost of proper processing, and the availability of informal dumping or low-quality recycling.
This leads appropriators to generate e-waste while overlooking the environmental and health externalities.
In this context, the goal of the governance structure is to reduce the stock of e-waste and ensure that the costs of proper disposal are borne by the actors who generate it, rather than by downstream communities and environments.

\subsubsection{Global E-waste Trade and the Basel Convention}

While e-waste is generated globally, its movement follows a distinct geographic pattern, typically flowing from the industrial North to the Global South.
This flow is driven primarily by economic disparities; the cost of extracting materials safely in developed nations is often higher than the value of the recovered materials.
Conversely, in developing nations, low labor costs make labor-intensive manual dismantling profitable \cite{das_global_2022, shittu_global_2021}.

To combat the trans-boundary movement of e-waste and other hazardous waste, the Basel Convention was established in 1989 \cite{gailhofer_protocol_2023} (\S\ref{sec:e-waste}). It restricts such movements and has since been ratified by 191 nations, making it one of the most influential international policies governing global e-waste flows.

\subsubsection{Local E-waste Governance}
Global mandates like the Basel Convention establish the overarching rules, and nations and local municipalities form the essential collective-choice and operational tiers required to enforce them.
Many countries have established their own management systems for e-waste, including the EU's WEEE Directive \cite{borner_toward_2018}, the USA's state-level frameworks (e.g., California’s SB 20/50) \cite{noauthor_political_nodate}, and Brazil's National Solid Waste Policy (PNRS) \cite{pedro_constructed_2021}.
While other nations maintain some degree of e-waste management, there is no universal baseline.
Generally, Global North countries possess more developed regulatory frameworks than the Global South \cite{das_global_2022}.

\subsection{Open Source Software and Linux}
Free and open source software (FOSS) is foundational for the vast majority of systems we use today \cite{hoffmann_value_2024}.
The Linux operating system is perhaps one of the most famous examples of such, created in 1991 by Linus Torvalds \cite{bretthauer_open_2001}.
The Linux Foundation \cite{about_linux_foundation} reports that at the time of writing, there are 6,773 unique annual contributors actively creating commits, pull requests, and raising issues \cite{linux_foundation_contrib}.

The community's shared goal is to democratize code, maintain high performance and security properties, and support a vast variety of hardware.
The Linux operating system is able to significantly extend hardware lifetime, from enterprise servers to embedded processors \cite{linux_lifecycle_power}.

Linus Torvalds' own role in this community has been studied extensively; he acts as a leader, an arbiter, a mediator, but also an active contributor in technical discussions \cite{noni_governance_2011, van_wendel_de_joode_managing_2004}.
All communication and coordination is done via mailing lists.

Patches are sent by individual contributors to Maintainers, members of the Linux community who are in charge of a particular subsystem.
Maintainers are selected by the community; they tend to be people who have working experience with that particular subsystem.
Linux goes through a release cycle roughly every two to three months, during which Torvalds selects patches sent to him by Maintainers to be included in the next version \cite{linux_dev_process}.
This chain-of-trust allows the community to function even with the large size of the project -- by the time a patch makes it to Torvalds, it's generally believed to be a good addition to the kernel.

\subsubsection{Linux and Negative Commons} \label{sec:linuxnegcommons}
Contributing to the kernel carries with it positive social status within the broad software development community.
As a result, there is an individual incentive to get as many changes as possible merged into the main branch.
This individual incentive is in tension with the shared incentive to maintain high code quality to achieve the aforementioned community goals.
The open source nature of the project is ultimately what makes the code base a commons.
A positive commons lens on this situation would result in claiming that "high code quality" is the stock being managed by the community. 
However, this lens falls short -- contributors are not extracting any particular resource from the code base, they are inherently \emph{adding} to the system.

A negative commons lens is more directly applicable to this situation; in the same way that donors add pollutants to an environment, contributors add code issues into the code base.
Both actions can occur either maliciously, on-purpose, or inadvertently, by-accident.
Regardless, these bugs are often hard to find once included, and can have a serious negative impact on broad populations.
A governance structure is embedded into the community, tasked with reducing the number of bugs introduced into the code base.
For this reason, we claim that the Linux kernel is an example of an immaterial NCPR.

\subsubsection{Bad Faith Bans}
In 2021, all contributors who used a University of Minnesota (UMN) email address were banned from submitting patches into the kernel \cite{umn_statement}.
A group of researchers were found to be intentionally contributing bugs. 
In response, Linux leadership decided to retroactively revert all patches previously submitted from UMN \cite{greg_kroah-hartman_patch_2021}.

This is one of the largest-scale bans that have ever taken place in the Linux community, although smaller bans due to "bad faith" submissions happen much more frequently.
Oftentimes, these individuals are banned for repeatedly not listening to Maintainer feedback.

\subsubsection{Toxicity and Bugs}
Ferreira et al. \cite{ferreira_shut_2021} conduct a qualitative analysis on 1,545 emails regarding rejected changes -- they find that instances of incivility appear in approximately two-thirds of the emails.
The 2007 Kernel Summit \cite{corbet_ks2007_2007} emphasized the importance for Maintainers to set better examples in mailing lists.
Torvalds' own stance on the toxic nature of the community is maybe best exemplified in a 2013 exchange regarding patches submitted late during a release cycle \cite{linus_torvalds_re_2013}:
\begin{quote}
    Greg, the reason you get a lot of stable patches seems to be that you make it easy to act as a door-mat ... You may need to learn to shout at people.
\end{quote}

Another contributor, Sarah Sharp, expressed concern about these comments \cite{sarah_sharp_re_2013}:
\begin{quote}
    Is this what we need in order to get improve \emph{-stable}? Linus Torvalds is advocating for physical intimidation and violence.
\end{quote}

In response, Torvalds continued to justify the use of harsh language, crediting it as a means of clear communication when it comes to important design decisions:
\begin{quote}
    I can't just say "please don't do that", because people won't listen. I say "On the internet, nobody can hear you being subtle", and I mean it.
\end{quote}

Leadership views \emph{harsh} communication as being a part of \emph{honest} communication, and therefore a means of keeping the kernel bug free.
This exchange, alongside plenty of others \cite{corbet_kernel_2013}, ultimately led to the adoption of the Contributor Covenant Code of Conduct (CCCC) \cite{linux_cccc}, and the creation of the Code of Conduct Committee.

\section{Methods}
We use the governance structures around e-waste and the Linux kernel as two case studies of NCPR management systems found within computing.
To evaluate the applicability of positive commons design principles to these NCPRs, we employ a qualitative, comparative, case study approach.
First, we perform a survey of primary sources to determine how the governance structure attempts to manage stock.
We then conduct a deductive qualitative analysis, coding these documents against Ostrom's eight design principles to determine how closely each NCPR's architecture aligns with traditional positive commons governance.
Finally, we assess the overall efficacy of each system in managing its respective stock, allowing us to observe potential correlations between successful negative commons management and adherence to Ostrom's principles.

\subsection{Data Collection}

Data was derived from a survey of existing literature, including academic analysis of governance structures, technical field reports, email communications, and policy frameworks.

\subsubsection{E-waste Data}
To investigate the governance of electronic waste through an NCPR lens, we conducted a comprehensive literature review focusing on real-world policy implementations and regulatory frameworks.

We performed a systematic search using Google Scholar to identify relevant academic literature.
The search query utilized a combination of the following keywords: "E-Waste", "WEEE", "governance", "commons", and "Ostrom".
This specific combination was chosen to bridge the gap between general waste management literature and theoretical frameworks of polycentric governance and common-pool resources.

The initial search yielded 625 results.
To ensure our dataset accurately reflected local manifestations of governance, we applied a two-stage filtering process.
First, during a title and abstract screening, we excluded purely theoretical papers lacking geographical application, as well as articles focused solely on technological innovation or quantitative waste metrics without policy analysis.
Second, during a full-text review, we retained only documents that explicitly discussed regulatory policies, institutional arrangements, or interactions with the informal sector.
From this refined pool, we narrowed the selection down to a final corpus of 10 articles that best exemplified diverse, localized governance approaches across different stages of economic development.
We ensured this final selection represented a diverse set of economic development stages and regulatory environments; Table \ref{tab:geo-dist} details the geographic distribution of these case studies.
\begin{table}[t]
    \centering
    \begin{tabular}{l|r}
         Region(s) & Number of papers \\
         \hline
         United States & 1 \\
         Brazil & 2 \\
         Netherlands & 1 \\
         South Africa & 1 \\
         Ghana & 1 \\
         China & 1 \\
         Multi-regional & 3 \\
    \end{tabular}
    \caption{Geographic distribution of the WEEE-related papers reviewed.}
    \label{tab:geo-dist}
\end{table}

\subsubsection{Linux Data}
Most Linux communication is fully public, done via the Linux Kernel Mailing List (LKML).
This archive provides a very rich record of the communication between different members that resulted in particular policy or contribution process change, as well as the feedback that Maintainers provide for particular patches.

To better understand the actual day-to-day manifestations of particular policies and attitudes, we construct a dataset of LKML emails.
To narrow down the number of emails considered, we queried LKML for appearances of "Nacked-by:" tags (which appear within emails where Maintainers reject a particular patch), and chose to analyze the 60 most recent, unique threads. 
This filter was applied to focus on how maintainers make and justify decisions regarding what code shouldn't be allowed within the code base -- in other words, scenarios where contribution to the common pool is being purposefully restricted.
These 60 threads spanned from February to early December in 2025, capturing nearly a year's worth of rejected kernel patches.

Manual inspection reveals that seven of those threads were discussing closed patches which were rejected in the past.
We remove these threads from our final dataset, keeping only the threads which concern themselves with active maintenance of the kernel.

To code the remaining 53 threads, each thread was read by one of the authors and labeled with a short phrase identifying why a particular patch was being rejected.
Any instances of interpersonal grievances between thread participants were noted, alongside if any advice was offered by maintainers.

The full list of identified codes, alongside the frequency with which each code appeared within the LKML repository, is shown in Table \ref{tab:lkml-codes}.
The common codes were then combined to form themes (\S \ref{sec:lkml-themes}), seeking to explain how the kernel's governance policies locally manifest.
\begin{table}[t]
    \centering
    \begin{tabular}{l|r}
         Code & Occurrences \\
         \hline
         Suggested Fix & 22 \\
         Big Picture & 20 \\
         Bad Style & 14 \\
         Logical Issue & 13 \\
         Contested (Maintained) & 13 \\
         Bad Faith & 12 \\
         Long-term Concern & 8 \\
         ``Keep the NAK'' & 8 \\
         Uncivil/Rude & 5 \\
         Known Issue in Other Subsystem & 5 \\
         Bad Patch Format & 5 \\
         Shutdown & 5 \\
         Contested (Overturned) & 4 \\
         Multiple Maintainers Criticize & 4 \\
         ``Kernel Dev Identity'' & 3 \\
         Cite Rules & 2 \\
         Slippery Slope & 1 \\
         No Tests & 1 \\
    \end{tabular}
    \caption{Codes identified in evaluating LKML emails.}
    \label{tab:lkml-codes}
\end{table}

Alongside this repository of LKML threads, documents detailing the Code of Conduct Committee, the CCCC itself, Linux Weekly News articles (the primary news source covering the open source community), and summit proceedings summaries were used to construct a comprehensive history of management.
These sources supplement the analysis in \S\ref{sec:lkml-themes}, providing evidence for the constitutional policies enacted by the governing body. 

\section{Evaluating Ostrom's Principles}

\subsection{The E-waste NCPR}
\label{sec:e-waste}
In this section, we evaluate whether the e-waste NCPR aligns with each of Ostrom's eight principles.

\subsubsection{Clearly Defined Boundaries}
\label{ewaste_boundary}
The Basel Convention fundamentally relies on sharp jurisdictional boundaries, presupposing a system of distinct nation-states to govern the movement of materials across borders. It attempts to harden these boundaries through Article 1 and the 2022 "E-waste Amendments" (entries A1181 and Y49), which legally subject all transboundary e-waste to Prior Informed Consent (PIC) \cite{gailhofer_protocol_2023}.
However, while its territorial borders are distinct, the definitional boundary of the resource itself frequently dissolves in practice due to the ambiguity between "waste" and "used goods."
While the European Union reinforces this resource boundary by requiring proof of functionality for exports, the Global South faces a "porous border" reality where these definitional gaps are readily exploited \cite{borner_toward_2018, shittu_global_2021}. 
In the West Bank, geopolitical fragmentation creates a jurisdictional governance vacuum where e-waste hubs emerge simply because the territory lies outside effective national control \cite{davis_beyond_2019}. 
Similarly, in the U.S. (a non-party to the Basel Convention), market actors exploit definitional gap to classify exports as "digital bridge" philanthropy, effectively smuggling waste through the "repair loophole" \cite{noauthor_political_nodate}. 
Thus, while Basel provides clear de jure jurisdictional boundaries, the de facto definitional boundary of the e-waste resource is fluid and contested.

\subsubsection{Congruence Between Appropriation Rules and Local Conditions}
\label{ewaste_congruence}
The Basel Convention’s primary appropriation mechanism assumes a level of bureaucratic capacity that is incongruent with local realities.
Because the constitutional-level framework fails to scale down, it forces a reliance on local, operational-level governance to achieve any meaningful congruence.
A primary example is the requirement for "Environmentally Sound Management" (ESM) based on "Best Available Techniques" (BAT) in the Basel Convention's technical guidelines \cite{gailhofer_protocol_2023, ESM_toolkot}.
While establishing high standards, this mandate fails to account for technical and economic discrepancies across regions.
In China and Ghana, both ratified states of the Basel Convention, top-down attempts to impose this formal, high-cost model (e.g., via "eco-parks") failed because they ignored the low operating costs and high recovery skills of the informal sector, driving the trade back underground \cite{lora2016trouble, asibey_dialogues_2025}.
Conversely, small-scale "constructed governance" in Brazil did demonstrate high congruence.
Instead of relying on distant bureaucratic forms, rules regarding pricing and logistics were negotiated directly between cooperatives and private buyers, resulting in higher compliance than international mandates could achieve \cite{pedro_constructed_2021}.

\subsubsection{Collective Decision-making}
\label{ewaste_decision}
The Basel Convention limits the participation to the state level via the Conference of the Parties (COP) \cite{gailhofer_protocol_2023}, effectively excluding the informal workers who manage the bulk of global flows.
This exclusion is mirrored in South Africa, where the "WEEE Producer Forum" marginalized key stakeholders to protect industry interests, leading to legitimacy crises and paralysis \cite{lawhon_contesting_2012}.
In contrast, we found that effective collective decision-making is observed where governance is localized.
The Netherlands utilizes a "monitoring council" where producers and municipalities jointly evaluate the system, allowing for adaptive management \cite{borner_toward_2018}.

\subsubsection{Monitoring}
\label{ewaste_monitor}
The Basel Convention relies on Article 13 (national reporting), a "fire-alarm" system that is functionally broken, with over 50\% of parties failing to report harmonized statistics in 2013\cite{gailhofer_protocol_2023, guidance2019}.
This failure creates a global monitoring vacuum, filled in the U.S. by a patchwork of competing market certifications (R2 vs. e-Stewards) that offer inconsistent transparency \cite{noauthor_political_nodate}.
Even though substantial effort has been devoted to monitoring e-waste, including institutionalized and socialized monitoring mechanisms, significant challenges remain.
For example, the Netherlands maintains a centralized national registry that aggregates data across the entire e-waste chain \cite{borner_toward_2018}.
However, the particular characteristics of e-waste mean that data quality still suffers from major inaccuracies, which are compounded by the blurry boundaries described in \autoref{ewaste_boundary}.

\subsubsection{Graduated Sanctions}
\label{ewaste_sanctions}
The Basel Convention’s primary sanction (Article 9) is a binary instrument (Legal vs. Illegal) that lacks an enforcement arm \cite{gailhofer_protocol_2023}, rendering it ineffective in contested zones like the West Bank where burning of e-waste persists without penalization \cite{davis_beyond_2019}.

\subsubsection{Conflict-Resolution Mechanisms}
\label{ewaste_conflict}
The Basel Convention’s Implementation and Compliance Committee (ICC) offers a facilitation body, but it is designed for diplomatic disputes between states, not for the daily operational conflicts of the waste trade. 
Without accessible mechanisms at the international level, disagreements within local governance structures tend to produce fragmentation rather than resolution.
In South Africa, the original multi-stakeholder e-waste body (eWASA) was intended to serve as a unified governance arena.
However, distrust in the organization's leadership and participatory processes led a group of manufacturers to form a rival body, the SA WEEE Producers' Forum \cite{lawhon_contesting_2012}. 
Rather than resolving the underlying disagreement about collective versus competitive recycling strategies, the split divided resources, created confusion among participants, and delayed the governance transition.
A similar pattern emerged in the U.S., where disagreements during the EPA's development of the R2 recycling standard led BAN to withdraw and create the rival e-Stewards certification \cite{noauthor_political_nodate}.
The two standards promote incompatible approaches to e-waste exports, yet no shared mechanism exists to adjudicate between them.
In both cases, the absence of a trusted conflict-resolution process caused e-waste governance to splinter.

\subsubsection{Minimal Recognition of Rights to Organize}
\label{ewaste_organize}
The Basel Convention remains largely silent on informal sector rights, treating e-waste management as a technical state function \cite{das_global_2022, shittu_global_2021}.
This silence often enables national governments in places like Ghana and China to criminalize or displace self-organized scrap hubs under the guise of "modernization" or Nature-based Solutions (NbS), explicitly violating the right to organize \cite{lora2016trouble, shittu_global_2021}.
This criminalization drives the trade underground, stripping it of oversight and paradoxically fueling a surge in unregulated, illicit activity.

\subsubsection{Nested Enterprises}
\label{ewaste_nested}
The breakdown of the nested network is a primary structural failure of the global e-waste landscape under the institutional governance of the Basel Convention.
While the Basel Convention attempts to link international obligations to national action via Regional Centres (BCRCs), the chain frequently breaks at the national level \cite{gailhofer_protocol_2023, das_global_2022}.
In China, central mandates decouple from local realities because local governments lack the incentive to enforce them against a profitable informal sector \cite{lora2016trouble}.
Conversely, the Dutch model exemplifies successful nesting on a continental scale.
The European WEEE Directive (constitutional level) is tightly coupled with national law (collective-choice level) and municipal execution (operational level), supported by robust information exchange \cite{borner_toward_2018}.
This contrast suggests that the Basel Convention's design is theoretically nested but practically disconnected in regions lacking strong institutional transmission belts.

\subsection{The Linux NCPR}\label{sec:lkml-themes}
In this section, we evaluate whether the Linux NCPR aligns with each of Ostrom's eight principles.

\subsubsection{Clearly Defined Boundaries}
\label{linux_boundary}
The hierarchical organization of the kernel into subsystems defines boundaries within which Maintainers exert control.
Further, the patch review process itself acts to clearly define who is able to make a contribution into the code base.
This principle heavily aligns itself with the \emph{Bad Faith} code, denoting times Maintainers rejected patches due to the person making the contribution.
To give an example \cite{toke_hoiland-jorgensen_re_2025}, the following was coded as \emph{Bad Faith}: 
\begin{quote}
    The request\_irq() change in ath\_ahb\_probe() is literally the same change that you sent once and that we had to revert. Not taking any more of these, sorry.
\end{quote}
Given the supporting institutional structure and the prevalence of these codes, we conclude that there is more than sufficient evidence of clearly defined boundaries.

\subsubsection{Congruence Between Appropriation Rules and Local Conditions}
\label{linux_congruence}
Maintainers often set their own standards for what contributions are acceptable within each subsystem, as defined in the Contributor Covenant Code of Conduct (CCCC).
Highly critical systems often move slower, with stricter requirements for incremental changes.
The \emph{Big Picture} code most directly aligned with this principle, representing times a Maintainer rejected a patch due to it containing features which were out of scope for a particular subsystem.
For example \cite{jakub_kicinski_re_2025}:
\begin{quote}
    I don't understand why patch 1 is a step in that direction, and you seem incapable of explaining it. So please either follow my suggestion on how to proceed with patch 2 without patch 1 in current form.
\end{quote}
Again, we find sufficient evidence for this principle.

\subsubsection{Collective Decision-making}
\label{linux_decision}
Maintainers are chosen by the community \cite{maintainers_chosen}, which determines how subsystem-level rules are created and enforced.
Beyond that, the community has been able to form institution-level governance bodies, for instance, the CCCC.
The \emph{Contested (Overturned)} code also contributes to evidence for collective-choice arrangements, denoting times a Maintainer originally rejected a patch but was then convinced by the contributor to include it.
Further, the \emph{``Keep the NAK''} code denotes times a Maintainer specifically said to include their rejection of the patch if it is ever resubmitted, so they can get a chance to voice their opinion in the future decision.
Once more, we conclude that there is good evidence for this principle.

\subsubsection{Monitoring}
\label{linux_monitor}
Maintainers are monitors.
Maintainers must read every submitted patch, and no patch ever makes it into the main branch without their approval.
Given the exhaustive nature of this requirement, monitoring is one of the central premises used by the Linux NCPR.

\subsubsection{Graduated Sanctions}
\label{linux_sanctions}
The most frequent code, \emph{Suggested Fix}, denotes times a patch was rejected, however, the Maintainer recommended a fix for the contributor to implement to get it approved.
Most times, there is no consequence to submitting a bug.
However if a pattern of \emph{Bad Faith} submissions is found, the individual could be banned from further contributing to a subsystem.
As an example \cite{lorenzo_stoakes_re_2025}:
\begin{quote}
    You are purposefully engaging in malicious compliance here...I suggest you find another part of the kernel to work upon.
\end{quote}

In the most extreme cases (such as the UMN affair), entire organizations have been banned from contributing to any subsystem. 
The level of penalty depends upon the severity of the infraction, and therefore, we claim that this NCPR makes heavy use of graduated sanctions.

\subsubsection{Conflict-Resolution Mechanisms}
\label{linux_conflict}
The CCCC established the Code of Conduct Committee to resolve non-trivial problems, as they come up.
The rules of the CCCC are sometimes cited in emails (coded as \emph{Cite Rules}), to direct the conflict resolution process in a productive direction.
\cite{markus_elfring_re_2025} is an example of such, where a patch was improperly formatted, and after a terse, tense, exchange, the rules were invoked to settle the matter.
Due to the clear definitions of responsibilities, alongside the invocations of the rules themselves, we conclude that this principle is also central to the success of the governance structure.

\subsubsection{Minimal Recognition of Rights to Organize}
\label{linux_organize}
Linux subsystems have been further split into sub-subsystems, with their own Maintainers.
The hierarchical nature of the chain of trust, alongside its ability to be extended, then becomes concrete evidence for the right to organize.
Further, the adoption of the CCCC was a bottom-up process, providing evidence for institutional change spearheaded through the organization of individual contributors.

\subsubsection{Nested Enterprises}
\label{linux_nested}
When Maintainers accept patches, they are further reviewed by a "higher-level" organization of Maintainers, up the chain-of-trust.
Torvalds can reject patches submitted by any Maintainer.
The overall structure of this system lends itself to the definition of a nested enterprise.

\section{Discussion}
The Linux community is able to sustain a kernel which powers the top 500 supercomputers \cite{supercomputers} and approximately 63\% of servers worldwide \cite{linux_share}.
It is a highly trusted and robust system; the community demonstrates strong evidence of all eight principles in action.
The governance of e-waste operates at a comparable global scale: 191 nations are party to the Basel Convention \cite{gailhofer_protocol_2023}, which seeks to manage the estimated 62 million tons of e-waste produced worldwide in 2022 \cite{GlobalEwasteMonitor2024}.
Yet under this regime, e-waste governance shows significant gaps, particularly in the application of effective monitoring, graduated sanctions, and nested enterprises.
In this discussion, we first assess both NCPRs against Ostrom's eight principles to answer the three research questions we posed, then focus on the importance of clearly defined boundaries, graduated sanctions, and nested frameworks.

\subsection{NCPR and Ostrom's Eight Principles}

Here we return to the three research questions posed in \S\ref{sec:intro}:

\emph{RQ1 (institutional design).}
Our analysis reveals a divergence in how Ostrom's principles are operationalized at the institutional level.
While both the Linux and e-waste governance systems theoretically acknowledge these principles, their efficacy hinges entirely on structural integration.
The Linux kernel embeds Ostrom's principles directly into its technical architecture. Boundaries, monitoring, and nested hierarchies are inescapable realities of the patch review process (\S\ref{linux_boundary}, \S\ref{linux_monitor}, \S\ref{linux_nested}). Conversely, the Basel Convention’s institutional design treats these principles as nominal, diplomatic constructs. Without the effective localized mechanisms to enforce boundaries (\S\ref{ewaste_boundary}) or apply graduated sanctions (\S\ref{ewaste_sanctions}), the international framework remains aspirational. We conclude that institutional design for NCPRs succeeds only when Ostrom’s principles are structurally unavoidable rather than merely bureaucratic.

\emph{RQ2 (local manifestation).}
Regarding local manifestations, the evidence indicates that rigid, top-down enforcement is fundamentally hostile to NCPR governance.
In e-waste, adherence to Ostrom’s principles -- and the subsequent success in managing the negative stock -- only emerges when governance localizes. Bottom-up, constructed governance models in Brazil and the Netherlands achieve the local congruence and collective decision-making that international mandates fail to impose \cite{borner_toward_2018, pedro_constructed_2021}.
Similarly, Linux thrives precisely because its subsystem maintainers act as highly localized, autonomous governors rather than strict enforcers of a single central decree (\S\ref{sec:lkml-themes}).
Effective NCPR governance, therefore, demands rules that are adaptable to local contexts.

\emph{RQ3 (overall efficacy).}
Ultimately, we conclude that Ostrom’s eight principles can become a functional blueprint for the governance of negative commons.
Our findings demonstrate that the ability to suppress "pollutants" scales directly with a system’s holistic adherence to these principles.
This holds true not only across the material/immaterial divide but also internally within domains.
Within the e-waste domain, the localized, constructed models of Brazil and the Netherlands succeed by exhibiting the same decentralized, nested structure found in the Linux community, whereas the broader Basel regime struggles where that nesting breaks down (\S\ref{ewaste_nested}, \S\ref{broken_nest}).
This suggests that institutional mechanisms corresponding to Ostrom's principles consistently emerge among successful negative commons governance structures.

\subsection{The Importance of Boundaries and Sanctions}

A critical disparity lies in the definition of boundaries between the two case studies.
Within the Linux kernel, the boundary is a part of the fundamental infrastructure of the community.
Despite considering contributions from all sources, the "chain of trust" ensures no contribution enters the main branch without Maintainer approval.
In contrast, e-waste governance relies on porous legal definitions of "waste" versus "used goods" -- illicit flows often slip through.

Furthermore, the Linux model extensively utilizes graduated sanctions, ranging from code critiques to bans, allowing for nuanced enforcement.
The Basel Convention relies on a binary legal status (legal vs. illegal) without a global enforcement arm, often resulting in no effective sanction for violations.

\subsection{The Broken Nest} \label{broken_nest}
Linux succeeds through a nested structure; Maintainers connect individual contributors to top-level governance, ensuring rules align with local technical realities.
E-waste governance, however, suffers from a broken nested network.
International mandates often disconnect from the operational reality of the informal sector in the Global South.
By failing to integrate these stakeholders, the governance lacks that congruence, paradoxically driving the trade underground.

\subsection{Limitations and Future Work} \label{sec:limitations}
It would be disingenuous for us to claim that the two case studies featured in this paper are perfectly analogous to one another, and that one governance system is directly applicable to the other situation.
For instance, the fundamental material/immaterial difference between them presents a gap that must be bridged when attempting to compare them directly.
Nonetheless, we find that in conducting an analysis of vastly different situations, it becomes much easier to ask: \emph{why do we struggle with the management of one system, and not the other? What makes them different?}
Observing these differences, in an explicit manner, allows us to then refine our understanding down to specific policies and to imagine different futures.
After all, a sustainable future includes the proper management of all kinds of negative commons.

Yet treating these two NCPRs as independent points along the computing lifecycle is itself a simplification.
The lifecycle of a computer binds them together: hardware is acquired, software is run upon it, and hardware is eventually discarded.
A well-governed kernel extends hardware lifetime \cite{linux_lifecycle_power}, delaying the obsolescence that produces e-waste in the first place.
Reducing the immaterial negative stock of bugs is, in this sense, partly continuous with reducing the material negative stock of discarded hardware.
We note one important subtlety: the Linux NCPR -- as we have framed it -- concerns contributions to the kernel; the code issues that contributors add to the shared system (\S\ref{sec:linuxnegcommons}).
By contrast, the connection to hardware longevity operates through an entirely different mechanism: the continued use of Linux to keep aging machines in service.
Our analytical frame, which deliberately isolated each system in order to test the portability of Ostrom's principles, was not able to capture this coupling.
Future work might treat the computing lifecycle as an interconnected system of negative commons, asking how governance of the immaterial stock propagates downstream into the material one, and vice versa.

We further acknowledge that this study does not measure the total quantitative effect of the Linux ecosystem nor the Basel Convention.
Instead, we grounded our analysis in the presence of governing principles at the "constitutional" level of these nested networks.
Future work could move beyond this qualitative reading to quantify the degree to which governance structures implement Ostrom's principles, yielding quantitative evidence to inform negative commons policymaking.

Additionally, the choice of framework informs how we evaluate each case study.
Monnin \cite{monnin_negative_2021} had previously laid out a slightly different framework than Deron's \cite{deron_sprouting_2025}, splitting negative commons into ones we must \emph{live with}, \emph{live with differently}, and \emph{live without}.
The two case studies in this paper fit squarely into the first two categories -- however future work is necessary to analyze the third category of \textit{live without}, to understand if CPR principles remain applicable when the goal is the total cessation of a resource’s production, rather than its management.

\section{Conclusion}
We set out to determine whether Ostrom’s design principles, originally derived for preserving positive commons, are applicable to the management of negative common pool resources. 
Our comparative analysis suggests a meaningful correlation.
The successful mitigation of bugs within the Linux kernel mirrors the community's strict adherence to all eight design principles. 
Conversely, the struggles of the Basel Convention illustrate the fragility of governance structures that only partially adhere to these principles while significantly neglecting others.

Empirical evidence from the local manifestations of these structures confirms that Ostrom’s framework contributes to the successful management of NCPRs. 
Future policy frameworks should therefore prioritize these design principles when crafting institutional designs aimed at managing negative commons, moving beyond simple mandates.
Furthermore, we encourage the LIMITS community to identify additional examples of computing-related NCPR governance, comparatively analyzing their successes and failures to better inform negative-commons policymaking.

\section{Acknowledgments}
We would like to thank Kurtis Heimerl for his guidance in solidifying the research direction and providing valuable comments on the draft. We also thank Dawn Walker, Eli Blevis, and Pedro Reynolds-Cuellar for their reviews.

\bibliographystyle{ACM-Reference-Format}


\end{document}